\RequirePackage[2020-02-02]{latexrelease}
\documentclass[showpacs,showkeys,preprintnumbers]{revtex4}%
\usepackage{float,amstext,amsthm,slashed,epigraph}
\usepackage{amsfonts}
\usepackage{amsmath}
\usepackage{graphicx}
\usepackage{dcolumn}
\usepackage{bm}
\usepackage{textcomp}
\usepackage{amssymb}%
\begin{document}
\title{QUANTIZATION OF NON-ABELIAN YANG-MILLS
THEORIES }
\author{Walaa I.\ Eshraim}
\affiliation{New York University Abu Dhabi, Saadiyat Island, P.O. Box 129188, Abu Dhabi, U.A.E.}

\begin{abstract}
A non-Abelian theory of fermions interacting with gauge bosons, the constrained system, is studied. The equations of motion for a singular system are obtained as total differential equations in many variables. The integrability conditions are investigated and the set of equations of motion is integrable. The Senjanovic and the canonical methods are used to quantize the system, and the integration is taken over the canonical phase space coordinates.\\

\end{abstract}

\maketitle

\section{Introduction}

\indent The Yang-Mills theories are field theories that describe the behavior of pure vectorial gauge fields. According to the relativistic quantum theory of electromagnetism, the interaction between two electrically charged particles is mediated through the exchange of virtual photons. As known, all kinds of particles can be thought of as excitations of some field. In the case of photons, they are the quantum excitations of the electromagnetic field, which is a vectorial gauge field invariant under $U(1)$ Abelian transformations. Consequently, in a similar way, the weak and strong interactions are described by the so-called non-Abelian vectorial gauge fields. The basic ingredients for the theoretical treatment of the fundamental interactions among elementary particles are the properties and symmetries of Yang-Mills theories. A particular case od Abelian Yang-Mills theory is the study of the classical properties of the Relativistic Electrodynamics, which is a well known topic in literature \cite{Yang1, Yang2}. But there is few studies of the classical properties of non-Abelian Yang-Mills theories in the field theory as seen in Refs. \cite{non1, non2, non3}. In Ref. \cite{nonWIE}, we studied the dynamics properties of the non-Abelian Yang-Mills theories as a constrained dynamics system \cite{constrained} by using Hamilton-Jacobi formulation.\\   
\indent The generalized Hamiltonian dynamics describing systems with constraints was initiated by Dirac \cite{D1,D2} and is widely used in investigating theoretical models in contemporary elementary particle physics \cite{P3,P4}. The presence of constraints in such theories requires care when applying Dirac's method, especially when first-class constraints arise since the first-class constraints are generators of gauge transformations which leads to gauge freedom. Dirac showed that the algebra of Poisson brackets determines a division of constraints into two classes: so-called first-class and second-class constraints. The first-class constraints are those that have zero Poisson brackets with all other constraints in the subspace of phase space in which constraints hold; constraints which are not first-class are by definition second-class. Most physicists believe that this distinction is quite important not only in classical theories but also in quantum mechanics \cite{P3,P4}.\\
\indent In the case of unconstrained systems, the Hamilton-Jacobi theory provides a bridge between classical and quantum mechanics. The first study of the Hamilton-Jacobi equations for arbitrary first-order actions was initiated by Santilli \cite{Santili}. The quantization and construction of the functional integral for theories with first-class constraints in canonical gauge was given by Faddeev \cite{Fad1, Fad2, Fad3}. Faddeev's method is generalized by Senjanovic \cite{Sen} to the case when second-class constraints appear in theory. Moreover, Fradkin \cite{Frank} considered the quantization of bosonic theories with first- and second-class constraints and extension to include fermions in such gauges. Gitman and Tyutin \cite{P3} discussed the canonical quantization of singular theories and the Hamiltonian formalism of gauge theories in an arbitrary gauge. In the recent past, the Hamiltonian-Jacobi approach \cite{G1, G2, S1} has been developed to investigate constrained systems. The equivalent Lagrangian method \cite{G3,Esh-Eq} is used to obtain the set of Hamilton-Jacobi partial differential equations (HJPDE). In this approach, the distinction between first- and second-class constraints is unnecessary. The equations of motion are written as total differential equations in many variables, which require investigating integrability conditions. Moreover, it is shown that gauge fixing, which is an essential procedure to study singular systems by Dirac's method, is not necessary if the canonical method is used \cite{S1}. The path integral formulation based on the canonical method obtained in Refs. \cite{Musl1, Musl2, Musl3, Musl4}.\\
\indent The non-Abelian Yang-Mill's theory is a constraint system. It has no genuine free limit because it usually has the three and the four-point gauge vertices at any non-zero value of the coupling however arbitrarily small. That is consistent with the so-called "background gauge field quantization" which seems to have fluctuations about classical space-time independent gauge fields. Consequently, it can't be gauged to zero at will since they come into the gauge invariant quantities which have non-trivial factors of structure constants in them which are fixed by the choice of the gauge group, and hence nothing can remove them by any weak coupling limit. But this constrained system, the singular system, is treated by using Dirac's and Hamilton-Jacobi's methods as seen in Ref. \cite{nonWIE}. Both of these methods are obtained for the non-Abelian Yang-Mills theory the same equations of motion as total differential equations in many variables. In Dirac's approach, the total Hamiltonian is composed by adding the constraints multiplied by Lagrange multipliers to the canonical Hamiltonian. In order to derive the equations of motion, one needs to redefine these unknown multipliers in an arbitrary way. However, in the Hamiltonian-Jacobi approach \cite{G4, Esh0, Esh1, Esh2, Esh3, Esh4, Esh5}, there is no need to introduce Lagrange multipliers to the canonical Hamiltonian. In the Hamilton-Jacobi approach, it is unnecessary to distinguish between first- and second-class constraints, there is no need to introduce any gauge fixing conditions as in Dirac's approach. Both the consistency conditions and the integrability conditions lead to the same constraints. 
In the present paper, we expand our investigation of the non-Abelian theory of fermions interacting with gauge bosons as a constrained system. The path integral quantization based on Senjanovic and canonical methods is applied to quantize the system.\\
\indent This paper is organized as follows. In Sec. II we present the path integral formulations for the Senjanovic method and the canonical path integral quantization. In Sec. IV we quantize the non-Abelian theory by using the Senjanovic and canonical methods. In Sec. V we outline our conclusions.


\section{Path Integral Formulation}

\indent In this section, we briefly review the Senjanovic
method and the Hamilton-Jacobi method for studying the path
integral for constrained systems.

\subsection{Senjanovic Method}

\indent Consider a mechanical system with $\alpha$ first-class
constraints and $\beta$ second-class constraints.Let the
first-class constraints be called $\phi_{a}$, the second -class
constraints $\theta_{a}$, and the gauge conditions associated with
the first-class constraints $\chi_{a}$. Let the $\chi_{a}$ be
chosen in such a way that $\{\chi_{a},\chi_{b}\}=0$.\\
\indent Then the expression for the $S$-matrix element is \cite{Sen}
\begin{equation}\label{1}
\left<Out\mid S\mid In\right>=\int exp
\left[i\int_{-\infty}^{\infty} (p_i \dot{q_i} - H_0)\, dt \right]
\prod_t \, d\mu (q(t), p(t)),
\end{equation}
where $H_{0}$ is the Hamiltonian of the system and the measure of
integration is defined by
\begin{equation}\label{2}
d\mu (q,p) = \left( \prod_{a=1}^\alpha \delta(\chi_a) \delta
(\phi_a) \right) det|| \{ \chi_a, \phi_a\}||\times
\prod_{b=1}^\beta \delta(\theta_b) \, det ||\{ \theta_a,
\theta_b\}||^\frac{1}{2} \prod_{i=1}^n dp_i \, dq^i.
\end{equation}
\subsection{Canonical path integral quantization}
$\\$\indent One starts from singular Lagrangian  $L\equiv
L(q_{i},{\dot{q}}_{i},t),\> i=1,2,\ldots,n$, with the Hess matrix
\begin{equation}\label{3}
A_{ij}=\frac{\partial^{2}L}{{\partial{\dot{q}}_{i}}\>{\partial{\dot{q}}_{j}}}\>,\> \,\,\,\,\,\,\,\,\,j=1,2,\ldots,n
\end{equation}
of rank $(n-r)$, $r<n$. The generalized momenta $p_{i}$
corresponding to the generalized coordinates $q_{i}$ are defined
as
\begin{align}
p_{a}&= \frac{\partial{L}}{\partial{\dot{q}}_{a}},\qquad {a =
1,2,\ldots,n-r}, \label{4}\\
p_{\mu}&= \frac{\partial{L}}{\partial{\dot{x}}_{\mu}},\qquad {\mu
= n-r+1,\ldots,n}.\label{5}
\end{align}
where $q_{i}$ are divided into two sets, $q_{a}$ and $x_{\mu}$.
Since the rank of the Hessian matrix is $(n-r)$, one may solve
Eq.(4) for ${\dot{q}}_{a}$ as
\begin{equation}\label{6}
{\dot{q}}_{a}={\dot{q}}_{a}(q_{i},{\dot{x}}_{\mu},p_{a};t).
\end{equation}
Substituting Eq. (6), into Eq. (5), we get
\begin{equation}\label{7}
p_{\mu}=-H_{\mu}(q_{i},{\dot{x}}_{\mu},p_{a};t).
\end{equation}
 The canonical Hamiltonian $H_{0}$ reads
\begin{equation}\label{8}
H_{0}=-L(q_{i},{\dot{x}}_{\nu},{\dot{q}}_{a};t)+
p_{a}{\dot{q}}_{a}-{\dot{x}}_{\mu}H_{\mu},\qquad
{\nu=1,2,\ldots,r}.
\end{equation}
The set of Hamilton-Jacobi Partial Differential Equations (HJPDE) is
expressed as
\begin{equation}\label{9}
H'_{\alpha}\bigg(x_{\beta},~q_{\alpha},\frac{\partial S}{\partial
q_{\alpha}} ,\frac{\partial S}{\partial
x_{\beta}}\bigg)=0,\qquad{\alpha,\beta=0,1,\ldots,r},
\end{equation}
where
\begin{equation}\label{10}
H'_{0}= p_{0}+H_{0}~,
\end{equation}
\begin{equation}\label{11}
 H'_{\mu}= p_{\mu}+H_{\mu}~.
\end{equation}
We define $p_{\beta}=\partial S[q_{a};x_{a}]/\partial x_{\beta}$
and $p_{a}=\partial S[q_{a};x_{a}]/\partial q_{a}$ with $x_{0}=t$
and $S$ being the action.\\
\indent Now the total differential equations are given as
\begin{align}
dq_{a}&=\,\frac{\partial H'_{\alpha}}{\partial p_{a}}\,dx_{\alpha},\label{12}\\
dp_{a}&=\frac{\partial H'_{\alpha}}{\partial q_{a}}\,dx_{\alpha},\label{13}\\
dp_{\beta}&=\frac{\partial H'_{\alpha}}{\partial
t_{\beta}}\,dx_{\alpha}, \label{14}
\end{align}
\begin{equation}\label{15}
dz= \bigg(-H_{\alpha}+p_{a}\frac{\partial H'_{\alpha}}{\partial
p_{a}}\bigg) dx_{\alpha},
\end{equation}
where $Z=S(x_{\alpha},q_{a})$. These equations are integrable if
and only if \cite{S1}
\begin{equation}\label{16}
dH'_{0}=0,
\end{equation}
\begin{equation}\label{17}
dH'_{\mu}=0,\,\;\quad \quad \mu=1,2,\ldots,r.
\end{equation}
If conditions (16) and (17) are not satisfied identically, one
considers them as a new constants and a gain consider their
variations. Thus, repeating this procedure, one may obtain a set
of constraints such that all variations vanish. Simultaneous
solutions of canonical equations with all these constraints
provide the set of canonical phase space coordinates
$(q_{a},p_{a})$ as functions of $t_{a}$; the canonical action
integral is obtained in terms of the canonical coordinates.
$H'_{\alpha}$ can be interpreted as the infinitesimal generator of
canonical transformations given by parameters $t_{\alpha}$,
respectively. In this case the path integral representation can be
written as \cite{Musl1, Musl3}.
\begin{equation}\label{18}
\left<Out\mid S\mid In\right>  =\int\prod_{a=1}^{n-p} dq^{a}
dp^{a} \exp{\left[ i\int_{t_{\alpha}}^{t'_{\alpha}}{\left(
-H_{\alpha}+ p_{a}\frac{\partial H'_{\alpha}}{\partial p_{a}}
\right)} dt_{\alpha}\right]},
\end{equation}
\indent $\,\,\,\quad \quad \quad \quad a=1,,\ldots,n-p,\quad \quad
\alpha=0,n-p+1,\ldots,n$.\\
\indent In fact, this path integral is an integration over the
canonical phase space coordinates $(q^{a},p^{a})$.

\section{Path integral quantization of non-Abelian Yang-Mills theories}
\indent Consider the Lagrangian density for
a non-Abelian theory of fermions interacting with gauge bosons as
\begin{equation}\label{19}
L=-\frac{1}{4}\>(F^{a}_{\mu\nu})^2+
\overline{\psi}(i\gamma^{\mu}D_{\mu}-m)\psi+\frac{1}{2\xi}(\partial^{\mu}
A^{a}_{\mu})^{2},
\end{equation}
where $\xi$ can be any finite constant and the covariant is given by\\
\begin{equation}\label{20}
D_\mu \psi(x)= (\partial_\mu-ig A^a_\mu(x)\,t^a )\, \psi(x)\,,
\end{equation}
where the $t^a$ are the generators of the gauge group $G$ written in the appropriate representation. The non abelian field strength is
\begin{equation}\label{21}
F^{a}_{\mu\nu}=\partial_{\mu}A^{a}_{\nu}-\partial_{\nu}A^{a}_{\mu}+gf^{abc}A^{b}_{\mu}A^{c}_{\nu},
\end{equation}
where $f^{abc}$ are the structure constants of the lie algebra and
$g$ represents the coupling constant.\\
The generalized momenta Eqs. (4) and (5) are
\begin{equation}\label{22}
\pi^{i}_{a}=\frac{\partial L}{\partial
{\dot{A}}^{a}_{i}}=-F^{0i}_{a},
\end{equation}
\begin{equation}\label{23}
\pi^{0}_{a}= \frac{\partial
L}{\partial{\dot{A}}^{a}_{0}}=\frac{1}{\xi}\>\partial^{\mu}
A^{a}_{\mu}\>,
\end{equation}
\begin{equation}\label{24}
p_{\psi}= \frac{\partial L}{\partial
\dot{\psi}}=i\>\overline{\psi}\gamma^{0}=-H_{\psi}\>,
\end{equation}
\begin{equation}\label{25}
p_{\overline{\psi}}= \frac{\partial L}{\partial
\dot{\overline{\psi}}}=0=-H_{\overline{\psi}}\>,
\end{equation}
\begin{equation}\label{26}
p_{\mu}= \frac{\partial L}{\partial \dot{A}_{\mu}}=0=-H_{\mu}\>.
\end{equation}
\indent Equations (22) and (23), respectively leads us to express
the velocities
\begin{equation}\label{27}
{\dot{A}}^{i}_{a}=\pi^{a}_{i}-\partial_{i}A^{0}_{a}+gf_{abc}\,A^{0}_{b}A^{i}_{c}\,,
\end{equation}
\begin{equation}\label{28}
{\dot{A}}^{0}_{a}=\xi\>\pi^{0}_{a}-\partial_{i}A^{a}_{i}.
\end{equation}
The Hamiltonian density is given by
\begin{align}\label{29}
H_{0}=\frac{1}{2}\>\pi^{a}_{i}\pi^{a}_{i}-\pi^{a}_{i}\,\partial_{i}A^{a}_{0}-gf^{abc}\pi^{i}_{a}A^{b}_{0}A^{c}_{i}+\frac{1}{2}\,\xi\>\pi^{a}_{0}\pi^{0}_{a}
-\pi^{a}_{0}\,\partial_{i}A^{a}_{i}+\frac{1}{4}\>F^{ij}_{a}F^{a}_{ij}-\overline{\psi}(i\gamma^{i}\partial_{i}+e\gamma^{\mu}A_{\mu}-m)\psi.
\end{align}
Making use of (9-11), we find for the set of HJPDE
\begin{equation}\label{30}
H'_{0}=\pi^{a}_{4}+H_{0}=0,
\end{equation}
\begin{equation}\label{31}
H'_{\psi}=p_{\psi}+H_{\psi}=p_{\psi}-i\overline{\psi}\,\gamma^0=0,
\end{equation}
\begin{equation}\label{32}
H'_{\overline{\psi}}=p_{\overline{\psi}}+H_{\overline{\psi}}=p_{\overline{\psi}}=0,
\end{equation}
\begin{equation}\label{33}
H'_{\mu}=p_{\mu}+H_{\mu}=p_{\mu}=0.
\end{equation}
\indent The equations of motion are obtained as total differential
equations follows:
\begin{align}
 dA^{i}_{a}&=\frac{\partial
H'_{0}}{\partial\pi^{a}_{i}}\>dt+\frac{\partial
H'_{\psi}}{\partial\pi^{a}_{i}}\>d\psi+\frac{\partial
H'_{\overline{\psi}}}{\partial\pi^{a}_{i}}\>d\overline{\psi}+\frac{\partial
H'_{\mu}}{\partial\pi^{a}_{i}}\>dA_{\mu}, \nonumber\\
&=[\pi^{a}_{i}-\partial_{i}A^{a}_{0}+gf^{abc}A^{b}_{0}\,A^{c}_{i}]\,dt,\quad
\quad \quad \quad \quad \quad \quad \quad \quad \,\,\,\,\label{34}
\end{align}
\begin{align}
\quad \quad \quad \quad dA^{0}_{a}&=\frac{\partial
H'_{0}}{\partial\pi^{a}_{0}}\>dt+\frac{\partial
H'_{\psi}}{\partial\pi^{a}_{0}}\>d\psi+\frac{\partial
H'_{\overline{\psi}}}{\partial\pi^{a}_{0}}\>d\overline{\psi}+\frac{\partial
H'_{\mu}}{\partial\pi^{a}_{0}}\>dA_{\mu},\nonumber\\
&=[\xi\,\pi^{a}_{0}-\partial^{i}A^{a}_{i}]\,dt,\quad \quad \quad
\quad \quad \quad \quad \quad \quad \quad \quad \quad \quad \quad
\,\,\,\,\,\label{35}
\end{align}
\begin{align}
 d\pi^{i}_{a}&=-\frac{\partial
H'_{0}}{\partial A^{a}_{i}}\>dt-\frac{\partial H'_{\psi}}{\partial
A^{a}_{i}}\>d\psi-\frac{\partial H'_{\overline{\psi}}}{\partial
A^{a}_{i}}\>d\overline{\psi}-\frac{\partial
H'_{\mu}}{\partial A^{a}_{i}}\>dA_{\mu},\nonumber\\
&=[gf^{abc}\pi^{i}_{c}A^{b}_{0}-\partial_{l}(F^{li}_{a}+\pi^{a}_{0})-F^{il}_{a}gf^{abc}A^{b}_{c}]\,dt,\quad
\quad \quad \,\,\,\label{36}
\end{align}
\begin{align}
 d\pi^{0}_{a}&=-\frac{\partial
H'_{0}}{\partial A^{a}_{0}}\>dt-\frac{\partial H'_{\psi}}{\partial
A^{a}_{0}}\>d\psi-\frac{\partial H'_{\overline{\psi}}}{\partial
A^{a}_{0}}\>d\overline{\psi}-\frac{\partial
H'_{\mu}}{\partial A^{a}_{0}}\>dA_{\mu},\nonumber\\
&=[\partial_{i}\pi^{i}_{a}+gf^{abc}\,\pi^{i}_{b}A^{c}_{i}]\,dt,\quad
\quad \quad  \quad \quad \quad \quad \quad \quad \quad
 \quad \quad \,\,\,\label{37}
\end{align}
\begin{align}
 dp_{\psi}&=-\frac{\partial H'_{0}}{\partial
\psi}\>dt-\frac{\partial H'_{\psi}}{\partial
\psi}\>d\psi-\frac{\partial H'_{\overline{\psi}}}{\partial
\psi}\>d\overline{\psi}-\frac{\partial
H'_{\mu}}{\partial \psi}\>dA_{\mu},\nonumber\\
&=[-\overline{\psi}\,(i\overleftarrow{\partial_{i}}\gamma^{i}-e\gamma^{\mu}\,A_{\mu}+m)]\,dt,
\quad \quad \quad \quad \quad \quad \quad \quad \quad \,\label{38}
\end{align}
\begin{align}
\quad \quad \quad \quad dp_{\overline{\psi}}&=-\frac{\partial
H'_{0}}{\partial \overline{\psi}}\>dt-\frac{\partial
H'_{\psi}}{\partial \overline{\psi}}\>d\psi-\frac{\partial
H'_{\overline{\psi}}}{\partial
\overline{\psi}}\>d\overline{\psi}-\frac{\partial
H'_{\mu}}{\partial \overline{\psi}}\>dA_{\mu},\nonumber\\
&=[(i\,\gamma^{i}\,\partial_{i}+e\gamma^{\mu}\,A_{\mu}-m)\psi]\,dt+i\,\gamma^{0}d\psi=0,
\quad \quad \quad \quad \,\label{39}
\end{align}
\begin{align}
\quad \quad \quad \quad dp_{\mu}&=-\frac{\partial H'_{0}}{\partial
A_{\mu}}\>dt-\frac{\partial H'_{\psi}}{\partial
A_{\mu}}\>d\psi-\frac{\partial H'_{\overline{\psi}}}{\partial
A_{\mu}}\>d\overline{\psi}-\frac{\partial
H'_{\mu}}{\partial A_{\mu}}\>dA_{\mu},\nonumber\\
&=(\overline{\psi}\,e\,\gamma^{\mu}\,\psi)\,dt=0,\quad \quad \quad
\quad \quad \quad \quad \quad \quad \quad \quad \quad
 \quad \quad \,\,\,\label{40}
\end{align}
\begin{equation}\label{41}
d\pi^{a}_{4}=-\frac{\partial H'_{0}}{\partial
t}\>dt-\frac{\partial H'_{\psi}}{\partial t}\>d\psi-\frac{\partial
H'_{\overline{\psi}}}{\partial t}\>d\overline{\psi}-\frac{\partial
H'_{\mu}}{\partial t}\>dA_{\mu}=0.
\end{equation}
\indent To check whether the set of Eqs. (34-41) is
integrable or not, we have to consider the total variations of the
constraints. In fact
\begin{equation}\label{42}
dH'_{\psi}=-\overline{\psi}\,(i\overleftarrow{\partial_{i}}\gamma^{i}-e\gamma^{\mu}\,A_{\mu}+m)\,dt-i\,d\overline{\psi}\,\gamma^{0}=0,
\end{equation}
\begin{equation}\label{43}
dH'_{\overline{\psi}}=(i\,\gamma^{i}\,\partial_{i}+e\gamma^{\mu}\,A_{\mu}-m)\psi\,dt+i\,\gamma^{0}\,d\psi=0.
\end{equation}
The constraints (31) and (32), lead us to obtain
$d\overline{\psi}$ and $d\psi$ in terms of $dt$
\begin{equation}\label{44}
d\overline{\psi}=[i\overline{\psi}\,(i\overleftarrow{\partial_{i}}\gamma^{i}-e\,\gamma^{\mu}\,A_{\mu}+m)\gamma^0]\,dt,
\end{equation}
\begin{equation}\label{45}
d\psi=[i\,\gamma^{0}(i\,\gamma^{i}\,\partial_{i}+e\,\gamma^{\mu}\,A_{\mu}-m)\psi]\,dt.
\end{equation}
The vanishing of the total differential of $H'_{\mu}$ leads to a
new constraint
\begin{equation}\label{46}
H''_{\mu}=\overline{\psi}\,e\,\gamma^{\mu}\,\psi,
\end{equation}
when we taking the total differential of $H''_{\mu}$, we show that
it's vanish
\begin{equation}\label{47}
dH''_{\mu}=0.
\end{equation}
Then the set of Eqs. (34-41) is integrable. Making use of
(15), (30), and (43), we can write the canonical action integral
as
\begin{align}\label{48}
 Z=\int
d^{3}x[2\pi^{i}_{a}\,\partial_{i}A^{a}_{0}+\frac{1}{2}\>\pi^{i}_{a}\pi^{i}_{a}-\frac{1}{4}\>F^{ij}_{a}F^{a}_{ij}+\frac{1}{2}\,\xi\>\pi^{a}_{0}\pi^{0}_{a}
+\overline{\psi}(i\gamma^{i}\partial_{i}+e\gamma^{\mu}A_{\mu}-m)\psi+(i\,\gamma^{0}\,\overline{\psi}+p_\psi)\dot{\psi}].
\end{align}
Now the S-matrix element which is define in Eq.(18) given as
\begin{align}\label{49}
\left<out|S|In\right> = \int \prod_{i,a}\,dA^{i}_{a}\,
d\pi^{i}_{a}\, d\psi\, d\overline{\psi}\,dp_{\psi}\>exp\,\,i\int
\bigg[2\pi^{i}_{a}\,\partial_{i}A^{a}_{0}+\frac{1}{2}\>\pi^{i}_{a}\pi^{i}_{a}+\frac{1}{2}\,\xi\>\pi^{a}_{0}\pi^{0}_{a}-\frac{1}{4}\>F^{ij}_{a}F^{a}_{ij}
\nonumber\\ \,\,\,\,\,\,\,\quad+\overline{\psi}(i\gamma^{i}\partial_{i}+e\gamma^{\mu}A_{\mu}-m)\psi+(i\,\gamma^{0}\,\overline{\psi}+p_\psi)\dot{\psi}\bigg]d^{4}x.
\end{align}
Now we will apply the Senjanovic method to quantize the non-Abelian Yang-Mills theories.
The total Hamiltonian is given as

\begin{align}\label{50}
H_{T}=\frac{1}{2}\>\pi^{a}_{i}\pi^{a}_{i}-\pi^{a}_{i}\,\partial_{i}A^{a}_{0}-gf^{abc}\pi^{i}_{a}A^{b}_{0}A^{c}_{i}+\frac{1}{2}\,\xi\>\pi^{a}_{0}\pi^{0}_{a}
-\pi^{a}_{0}\,\partial_{i}A^{a}_{i}+\frac{1}{4}\>F^{ij}_{a}F^{a}_{ij}\nonumber\\-\overline{\psi}(i\gamma^{i}\partial_{i}+e\gamma^{\mu}A_{\mu}-m)\psi
+\lambda_{\psi}(p_{\psi}-i\gamma^{0}\,\overline{\psi})+\lambda_{\overline{\psi}}\,p_{\overline{\psi}}+\lambda_{\mu}\,p_{\mu},
\end{align}
where $\lambda_{\psi}, \lambda_{\overline{\psi}}$ and
$\lambda_{\mu}$ are a Lagrange multipliers to be determined. From
the consistency conditions, the time derivative of the primary
constraints should be zero, that is
\begin{equation}\label{51}
{\dot
H}'_{\psi}=\{H'_{\psi},H_{T}\}=-\overline{\psi}\,(i\overleftarrow{\partial_{i}}\gamma^{i}-e\,\gamma^{\mu}\,A_{\mu}+m)-i\,\lambda_{\overline{\psi}}\,\gamma^{0}\approx0,
\end{equation}
\begin{equation}\label{52}
{\dot
H}'_{\overline{\psi}}=\{H'_{\overline{\psi}},H_{T}\}=(i\,\gamma^{i}\,\partial_{i}+e\gamma^{\mu}\,A_{\mu}-m)\psi+i\,\gamma^{0}\,\lambda_{\psi}\approx0,
\end{equation}
\begin{equation}\label{53}
{\dot
H}'_{\mu}=\{H'_{\mu},H_{T}\}=\overline{\psi}\,e\,\gamma^{\mu}\,\psi\approx0.
\end{equation}
The Eqs.(51) and (52) fix the multipliers
$\lambda_{\overline{\psi}}$ and  $\lambda_{\psi}$ respectively as
\begin{equation}\label{54}
\lambda_{\overline{\psi}}=i\,\overline{\psi}\,(i\overleftarrow{\partial_{i}}\gamma^{i}-e\,\gamma^{\mu}\,A_{\mu}+m)\gamma^{0},
\end{equation}
\begin{equation}\label{55}
\lambda_{\psi}=i\,\gamma^{0}(i\,\gamma^{i}\,\partial_{i}+e\gamma^{\mu}\,A_{\mu}-m)\psi.
\end{equation}
Eq. (53) lead to the secondary constraints
\begin{equation}\label{56}
H''_{\mu}=\overline{\psi}\,e\,\gamma^{\mu}\,\psi\approx0.
\end{equation}
There are no tertiary constraints, since
\begin{equation}\label{57}
{\dot H}''_{\mu}=\{H''_{\mu}, H_{T}\}=0.
\end{equation}
By taking suitable linear combinations of constraints, one has to
find the first-class, that is
\begin{equation}\label{58}
\Phi_{1}= H'_{\mu}= p_{\mu},
\end{equation}
whereas the constraints
\begin{equation}\label{59}
\Phi_{2}= H'_{\psi}= p_{\psi}-i\,\gamma^{0}\,\overline{\psi},
\end{equation}
\begin{equation}\label{60}
\Phi_{3}= H'_{\overline{\psi}}= p_{\overline{\psi}},
\end{equation}
\begin{equation}\label{61}
\Phi_{4}=H''_{\mu}=\overline{\psi}\,e\,\gamma^{\mu}\,\psi=0
\end{equation}
are second-class.\\
The equations of motion are read as
\begin{equation}\label{62}
{\dot
A}^{a}_{0}=\{A^{a}_{0},H_{T}\}=\xi\,\pi^{a}_{0}-\partial^{i}A^{a}_{i}\>,
\end{equation}
\begin{equation}\label{63}
{\dot
A}^{a}_{i}=\{A^{a}_{i},H_{T}\}=\pi^{a}_{i}-\partial_{i}A^{a}_{0}+gf^{abc}A^{b}_{0}\,A^{c}_{i}\>,
\end{equation}
\begin{equation}\label{64}
{\dot \psi}=\{\psi,H_{T}\}=\lambda_{\psi}\>,
\end{equation}
\begin{equation}\label{65}
{\dot
{\overline{\psi}}}=\{\overline{\psi},H_{T}\}=\lambda_{\overline{\psi}}\>,
\end{equation}
\begin{equation}\label{66}
{\dot A}_{\mu}=\{A_{\mu},H_{T}\}=\lambda_{\mu}\>,
\end{equation}
\begin{equation}\label{67}
{\dot \pi}^{0}_{a}=\{\pi^{0}_{a},H_{T}\}
=\partial_{i}\pi^{i}_{a}+gf^{abc}\,\pi^{i}_{b}A^{c}_{i}\>,
\end{equation}
\begin{equation}\label{68}
{\dot \pi}^{i}_{a}=\{\pi^{i}_{a},H_{T}\}
=gf^{abc}\pi^{i}_{c}A^{b}_{0}-\partial_{l}(F^{li}_{a}+\pi^{a}_{0})-F^{il}_{a}gf^{abc}A^{b}_{c}\>,
\end{equation}
\begin{equation}\label{69}
{\dot p}_{\psi}=\{p_{\psi},
H_{T}\}=-\overline{\psi}\,(i\overleftarrow{\partial_{i}}\gamma^{i}-e\gamma^{\mu}\,A_{\mu}+m),
\end{equation}
\begin{equation}\label{70}
{\dot p}_{\overline{\psi}}=\{p_{\overline{\psi}},
H_{T}\}=(i\,\gamma^{i}\,\partial_{i}+e\gamma^{\mu}\,A_{\mu}-m)\psi+i\,\gamma^{0}\lambda_{\psi}\>,
\end{equation}
\begin{equation}\label{71}
{\dot p}_{\mu}=\{p_{\mu},
H_{T}\}=\overline{\psi}\,e\,\gamma^{\mu}\,\psi.
\end{equation}
We will contact ourselves with a partial gauge fixing by
introducing gauge constraints for the first-class primary
constraints only, just to fix the multiplier $\lambda_{\mu}$ in
Eq. (50). Since $p_{\mu}$ is vanishing weakly, a gauge choice near
at hand would be:
\begin{equation}\label{72}
\phi'_{1}=A_{\mu}=0.
\end{equation}
But for this forbids dynamics at all, since the requirement
${\dot A}_{\mu}=0$ implies $\lambda_{\mu}=0$.\\
\indent Making use of Eq.(1), we obtain

\begin{align}\label{73}
\left<out|S|In\right> =N \int \prod DA^{i} exp\bigg[i\{\int
d^{4}x\bigg(\frac{1}{2}(\partial_{i}A^{0}+{\dot{A}}^i)^2
+\frac{1}{2}\>m^{2}(A_{0}A^{0}+A_{i}A^{i})-\frac{1}{4}\>F^{ij}F_{ij}
-A_{0}J^{0}-A_{i}J^{i}\bigg)\} \bigg],
\end{align}

\begin{align}\label{74}
\left<out|S|In\right> =N \int \prod DA^{i} exp\bigg[i\bigg\{\int
d^{4}x\bigg(\frac{1}{2}\>{\dot{A}}^2-\frac{1}{4}\>\>F^{ij}F_{ij}+J\cdot
A+\frac{1}{2}\>m^{2}(A_{0}A^{0}+A_{i}A^{i})\nonumber\\-\frac{1}{2}\>A^0(2\>\triangledown\cdot
\dot A+2\>J_0+ \triangledown ^{2}A_{0})\bigg)\bigg\} \bigg],
\end{align}

where $N$ is a constant.\\
Eq. (54) can be witten as
\begin{align}\label{75}
\left<out|S|In\right> =N \int \prod DA^{i} exp\bigg[i\bigg\{\int
d^{4}x\bigg(\frac{1}{2}\>{\dot{A}}^2-\frac{1}{4}\>\>F^{ij}F_{ij}+J\cdot
A+\frac{1}{2}\>m^{2}(\phi^2+A_{i}A^{i})\nonumber\\-\frac{1}{2}\>\phi\,(2\>\triangledown\cdot
\dot A+2\>\rho+ \triangledown ^{2}\phi)\bigg)\bigg\} \bigg],
\end{align}
Equations (46) and (48) lead us to obtain
\begin{equation}\label{76}
\triangledown\cdot \dot A=-\triangledown^2\>\phi+m^{2}\phi-\rho\>.
\end{equation}
Inserting eq.(56) in the integral (55) gives
\begin{align}\label{77}
\left<out|S|In\right> =N \int \prod DA^{i} exp\bigg[i\bigg\{\int
d^{4}x\bigg(\frac{1}{2}\>{\dot{A}}^2-\frac{1}{4}\>\>F^{ij}F_{ij}+J\cdot
A+\frac{1}{2}\,\phi\>\triangledown^{2}\phi
-\frac{1}{2}\>m^{2}(\phi^2-A_{i}A^{i})\bigg)\bigg\} \bigg].
\end{align}
One should notice that the term
$\frac{1}{2}\>{\dot{A}}^2-\frac{1}{4}\>\>F^{ij}F_{ij}+J\cdot A$
leads to a solution for $A$, in which
\begin{equation}\label{78}
A=\int G\>\eta\> d^{3}x.
\end{equation}
Thus $G$ has an inverse.\\

\section {Conclusion}

In this paper, we have investigated, a constrained system, the non-Abelian Yang-Mills theories by using Senjanovic and canonical methods to quantize the system.
We have obtained the equations
of motion of this system by Dirac’s and Hamilton-Jacobi method. In the Dirac method the total Hamiltonian
composed by adding the constraints multiplied by Lagrange multipliers to the canonical Hamiltonian. In order to
drive the equations of motion, one needs to redefine these unknown multipliers in an arbitrary way. However, in the
Hamilton-Jacobi formalism, there is no need to introduce Lagrange multipliers to the canonical Hamiltonian. Both the
consistency conditions and integrability conditions lead to the same constraints. In the Hamilton-Jacobi formulation,
the equations of motion are obtained directly by using HJPDES as total differential equations.
Path integral quantization of the non-Abelian theories is obtained by using the Senjanovic method and the canonical path integral quantization. Both methods give the same results. However, gauge fixing is not necessary
to obtain the path integral formulation for field theories if the
canonical formulation is used. Since this system is integrable,
$H'_{0}$ and $H'$ can be interpreted as infinitesimal generators
of canonical transformations given by parameters $t$ and $A^0$,
respectively. In this case the path integral is obtained as an
integration over the canonical phase-space coordinates $A^{i},
\pi^{i}$. Faddeev and Popov treatments \cite{P3, P4} need gauge-fixing
conditions to obtain the path integral over the canonical
variable.

\section*{Acknowledgments}

The authors acknowledge support from the NYUAD Center for Interacting Urban Networks (CITIES) through Tamkeen under the NYUAD Research Institute Award CG001.

\end{document}